\documentclass[runningheads]{llncs}
\usepackage{slashbox}
\usepackage{romannum}
\usepackage{amssymb,mathrsfs,amsfonts}
\usepackage{physics}
\usepackage{subcaption}
\usepackage{amsmath}
\usepackage{algorithm}
\usepackage[noend]{algpseudocode}
\usepackage[bottom]{footmisc}
\usepackage{tikz}
\usepackage{graphicx}
\usepackage{grffile}
\usepackage{amsmath}
\newcommand{\inlineeqnum}{\refstepcounter{equation}~~\mbox{(\theequation)}}
\usepackage{amsmath}
\usepackage{upgreek}
\usepackage{cite}
\usepackage{float}
\usepackage{url}
\usepackage{enumitem}
\usepackage{booktabs}
\usepackage{algorithm}
\usepackage[noend]{algpseudocode}
\usepackage[section]{placeins}
\usepackage{array}
\usepackage[bottom]{footmisc}
\usepackage{tikz}
\usetikzlibrary{patterns}
\usepackage{csvsimple}
\usepackage{graphicx}
\usepackage{caption}
\usepackage{url}
\usepackage[T1]{fontenc}
\usepackage{listings}
\usepackage{pgfplotstable}
\usepackage{pgfplots}
\pgfplotsset{compat=newest}
\usepackage{rotating}
\usepackage{tcolorbox}
\usepackage{bibnames}
\usepackage{booktabs}
\usepackage{lipsum}
\usepackage{subcaption}
\usepackage{arydshln}
\usepackage{stackengine}
\usetikzlibrary{automata,positioning}
\newcommand{\A}{{\cal A}}

\makeatletter
\def\BState{\State\hskip-\ALG@thistlm}
\makeatother

\usepackage[edges]{forest}

\usepackage{import}
\usetikzlibrary{quotes,arrows.meta}
\usetikzlibrary{positioning}

\def\edgecolor{rgb:blue,4;red,1;green,4;black,3}
\newcommand{\midarrow}{\tikz \draw[-Stealth,line width =0.8mm,draw=\edgecolor] (-0.3,0) -- ++(0.3,0);}

\usepackage{Ball}
\usepackage{Box}
\usepackage{RightBandedBox}

\usetikzlibrary{positioning}
\usetikzlibrary{3d} 

\def\ConvColor{rgb:yellow,5;red,2.5;white,5}

\def\PoolColor{rgb:red,1;black,0.3}

\def\FcReluColor{rgb:blue,5;red,5;white,4}
  
\def\OutColor{rgb:green,7;white,4}

\definecolor{foldercolor}{RGB}{124,166,198}

\tikzset{pics/folder/.style={code={%
    \node[inner sep=0pt, minimum size=#1](-foldericon){};
    \node[folder style, inner sep=0pt, minimum width=0.0*#1, minimum height=0.6*#1, above right, xshift=0.00*#1] at (-foldericon.west){};
    \node[folder style, inner sep=0pt, minimum size=#1] at (-foldericon.center){};}
    },
    pics/folder/.default={20pt},
    folder style/.style={draw=foldercolor!80!black,top color=foldercolor!40,bottom color=foldercolor}
}

\forestset{is file/.style={edge path'/.expanded={%
        ([xshift=\forestregister{folder indent}]!u.parent anchor) |- (.child anchor)},
        inner sep=0pt},
    this folder size/.style={edge path'/.expanded={%
        ([xshift=\forestregister{folder indent}]!u.parent anchor) |- (.child anchor) pic[solid]{folder=#1}}, inner xsep=0.6*#1},
    folder tree indent/.style={before computing xy={l=#1}},
    folder icons/.style={folder, this folder size=#1, folder tree indent=1*#1},
    folder icons/.default={12pt},
}

\usepackage{blindtext}

\usepackage[T1]{fontenc}
\usepackage[font=small,labelfont=bf,tableposition=top]{caption}

\DeclareCaptionLabelFormat{andtable}{#1~#2  \&  \tablename~\thetable}

\DeclareUrlCommand\ULurl{
  
  }

\usepackage{hyperref}

\begin{document}
\sloppy

\title{Complete Deep Computer-Vision Methodology for Investigating Hydrodynamic Instabilities}

\author{
Re'em Harel\inst{1,2}
\and Matan Rusanovsky\inst{2,3}
\and Yehonatan Fridman\inst{2,3}
\and Assaf Shimony\inst{4}
\and Gal Oren\inst{3,4}\thanks{Corresponding author}}

\authorrunning{R. Harel, M. Rusanovsky, Y. Fridman,  A. Shimony, G. Oren}
\titlerunning{Complete CVDL Methodology for Investigating Hydrodynamic Instabilities}

\institute{
Department of Physics, Bar-Ilan University, Ramat-Gan, Israel
\and Israel Atomic Energy Commission, P.O.B. 7061, Tel Aviv, Israel
\and Department of Computer Science, Ben-Gurion University of the Negev, P.O.B. 653, Be'er-Sheva, Israel\\
\and Department of Physics, Nuclear Research Center-Negev, P.O.B. 9001, Be'er-Sheva, Israel \\
\email{reemharel22@gmail.com, matanru@post.bgu.ac.il, fridyeh@post.bgu.ac.il, shimonya@gmail.com, orenw@post.bgu.ac.il}}

\date{}

\maketitle

\begin{abstract}
In fluid dynamics, one of the most important research fields is hydrodynamic instabilities and their evolution in different flow regimes. The investigation of said instabilities is concerned with the highly non-linear dynamics. Currently, three main methods are used for understanding of such phenomenon -- namely analytical models, experiments and simulations -- and all of them are primarily investigated and correlated using human expertise. In this work we claim and demonstrate that a major portion of this research effort could and should be analysed using recent breakthrough advancements in the field of Computer Vision with Deep Learning (CVDL, or Deep Computer-Vision). Specifically, we target and evaluate specific state-of-the-art techniques -- such as Image Retrieval, Template Matching, Parameters Regression and Spatiotemporal Prediction -- for the quantitative and qualitative benefits they provide. In order to do so we focus in this research on one of the most representative instabilities, the Rayleigh-Taylor one, simulate its behaviour and create an open-sourced state-of-the-art annotated database (\textit{RayleAI}). Finally, we use adjusted experimental results and novel physical loss methodologies to validate the correspondence of the predicted results to actual physical reality to prove the models efficiency. The techniques which were developed and proved in this work can be served as essential tools for physicists in the field of hydrodynamics for investigating a variety of physical systems, and also could be used via Transfer Learning to other instabilities research. A part of the techniques can be easily applied on already exist simulation results. All models as well as the data-set that was created for this work, are publicly available at: \ULurl{https://github.com/scientific-computing-nrcn/SimulAI}.
\end{abstract}


\section{Introduction}\label{introduction}

The Rayleigh-Taylor instability (RTI) occurs in an interface between two fluids with different densities in which the lighter fluid pushes the heavier fluid~\cite{sharp1983overview}. RTI is found in many hydrodynamic experiments and natural phenomena such as Inertial Confinement Fusion (ICF), water suspended on oil in earth's gravity, astrophysical systems and many more \cite{drazin2002introduction}. Numerous experiments studying the growth of the instability and its effects on other phenomena are performed constantly all over the world \cite{read1984experimental,dalziel1993rayleigh,dimonte1996turbulent,waddell2001experimental} due to its importance. Generally speaking, there are two types of experimental platforms for investigating the evolution of RTI: Liquid or gas systems (for example~\cite{read1984experimental,dalziel1993rayleigh,dimonte1996turbulent}) and High-Energy-Density Physics (HEDP) systems, in which the fluids are in plasma state after being heated by powerful lasers (for example \cite{knauer2000single,remington2019rayleigh}). In the former systems, it is difficult to control the initial perturbation but the time difference between consecutive frames is short comparing to the duration of the experiment, which typically can vary from milliseconds to seconds. Therefore, it is feasible to obtain with a fast camera tens or more frames per experimental shots. In the latter systems, the initial perturbation can be machined precisely prior to the laser drive, while the materials are in solid state, but the time scales are much shorter (about tens of nanoseconds) and only one or at most few frames can be obtained from a single experimental shot. Therefore, the experimental data from both types of experimental systems contain a partial reflection of the instability -- either the exact initial perturbation or the detailed temporal evolution of the instability is missing, while both are crucial for understanding the phenomenon. The growth of the perturbation in RTI depends on numerous variables such as viscosity, ablation, surface tension, small density gradients and more. These variables are of different importance in different physical and experimental systems. In this work we consider the case of two incompressible and immiscible fluids and a single-mode sinusoidal initial perturbation. In this case, the early growth is exponential in time and is given by (via linear stability theory): $e^{\sqrt{\A k g} t} \inlineeqnum\label{eq:linear}$ where $k$ is the wave number $k=\frac{2\pi}{\lambda}$, $\lambda$ is the wavelength, $g$ is the earth's gravity (and in the general case the acceleration of the system), and $\A$ is the well known Atwood number, given by $\A = \frac{\rho_1 - \rho_2}{\rho_1 + \rho_2} \inlineeqnum$ where $\rho_1$ is the density of heavier fluid and $\rho_2$ of the lighter one. In the late non-linear growth of such a single-mode perturbation, bubbles of the lighter fluid penetrate into the heavy fluid and spikes of heavy fluid penetrate into the light fluid in constant velocities, given by \cite{goncharov2002analytical}: $u_{b/s}=\sqrt{\frac{2Ag\lambda}{c_d\left(1\pm A\right)}} \inlineeqnum$ where $u_b$ and $u_s$ are the velocities of the bubble and the spike, respectively. $c_d$ is the drag coefficient, which equals to 6$\pi$ and 3$\pi$ for 2D and 3D, respectively. As the perturbation grows, the shear velocities on the sides of the bubbles and the spikes create vortices due to Kelvin-Helmholtz instability (KHI), in which the two materials mix in small length scales.

Needless to say that in reality (experiments), knowing the exact conditions of the density and the viscosity of the two fluids is unrealistic. A possible way to bridge this gap is via simulations~\cite{youngs1984numerical} (which are much cheaper than performing additional experiments): Given the correct initial parameters one can simulate the experiment and extract the missing time frames. However, initiating the simulation with the exact initial parameters is impossible due to the experimental uncertainties. Nevertheless, it is still a viable solution as one can run parameters-sweep and select the most similar simulation in comparison to the experiment. However, this solution might be difficult as there are many different parameters (both physical parameters with uncertainties and parameters in the analysis of the experimental results), which makes this process hard for a human.
In the past years, the usage of different CVDL techniques in this scientific area is growing \cite{spears2018deep, humbird2019transfer, gonoskov2019employing, avaria2019hard, humbird2019machine, gaffney2019making, kustowski2019transfer, kim2019machine, gonoskovemploying}, and as it progresses, the above problem might also be solved using CVDL, since the introduction of CVDL techniques to the CFDs field proved to yield excellent results \cite{raissi2019physics,raissi2019deep,mohan2018deep,wang2018model,lye2020deep,kutz2017deep,huang2019applications}. As different techniques in the CVDL field might be necessary in order to solve the problem, we define and devise several key problems, which collaboratively will enhance our understanding of RTI and other physical phenomena:
\begin{enumerate}
	\item Given a diagnostic \textit{image} from a simulation/experiment, \textit{sort} a database in accordance with an image similarity score to the input.
	\item Given a diagnostic \textit{image} from simulation/experiment, \textit{extract the parameters} of the simulation that yields a best match to an image in a database.
	\item Given a \textit{partial template} of a phenomena from a simulation/experiment, \textit{find} matches in a database and \textit{sort} them in accordance with an image similarity score to the input.
	\item Given a \textit{set of images} that correspond to a set of time steps, and a time parameter $T$, \textit{predict} future non-existing time steps images.
\end{enumerate}

\vspace{-0.7cm}

\begin{table}[H]
    \begin{center}
        \begin{tabular}{p{0.4cm}|p{3.5cm}|p{7.5cm}|}
            {\#} & \textbf{Task} & \textbf{Quantitative \& Qualitative Values} \\
            \hline
            \Romannum{1}. & Database sorting by an image similarity score. & $\bullet$ Meaningful order. \newline $\bullet$ Extraction of non-regressive parameters. \\
            \hline
            \Romannum{2}. & Regressive parameter extraction. & $\bullet$ Physical parameters that fit the experimental data. \newline $\bullet$ Uncertainty margins of the model training. \\
            \hline
            \Romannum{3}. & Find and sort partial templates by an image similarity score. & $\bullet$ Meaningful order. \newline $\bullet$ A more extensive and more reliable matching survey which results in a decrease of the uncertainty margins for the template assumption comparing to analysis of the full images only. \\
            \hline
            \Romannum{4}. & Temporal inter/extrapolation of experiments. & $\bullet$ Data augmentation for low-data experiments. \newline $\bullet$ Assurance and extension of a model. \\ 
            \hline
        \end{tabular}
        \caption{Quantitative and qualitative values for defined problems using CVDL.}
        \label{table:advantages}
    \end{center}
\end{table}

\vspace{-1cm}

The completion of the tasks above using CVDL techniques have both quantitative and qualitative values over classical optimization techniques, as presented in Table~\ref{table:advantages}. In the first technique, a database (in our case, images) is being sorted by an image similarity score. When the similarity score is being weighted according to the physical significance of the features in the data, the resulted order is meaningful. In addition, the physical simulation parameters that yield the maximal image similarity (for example, with the experimental results) can be used as a non-regressive optimization. Similarly, uncertainty margins for the physical parameters can be calculated by defining a minimal required similarity factor. 

The second CVDL technique is an advanced optimization method: Given experimental results and corresponding simulations with a set of free parameters, the technique provides the values of the free parameters for a best fit to the experimental results \textit{iff} the model training and validation loss convergence to approximately the same (small) value. The value of this method over the parameter extraction in the first technique is due to the regression process, which is significant when the simulation database is incomplete.

The third CVDL technique is similar to the first one, except for focusing on partial templates in the images instead of analyzing the whole image. It is valuable for physical problems in which there is a measurable pattern which is sensitive to the physical parameters. For example, in \cite{wan2015observation}, the evolution of vortices, created by supersonic KHI, was measured and compared with hydrodynamic simulations. The analysis in \cite{wan2015observation} was based on the large-scale structures (i.e. the widths of the vortices). The medium-scale structures (i.e. the roll-ups within the vortices) were measured in the experiment and were compared to the simulations qualitatively. A more detailed analysis of the templates of the roll-ups could provide additional physical insights such as the effect of viscosity in the experimental conditions. Another example, which is relevant to the evolution of RTI, is originated from morphology differences between experiments in HEDP platforms and simulations of them. A detailed analysis of the morphology of the bubbles and the spikes can provide insights on magnetic effects due to the plasma conditions in these experiments \cite{fryxell2010possible} or perhaps other physical effects. A third example is the measurement of ablative RTI \cite{kuranz2018high,huntington2018ablative} which is relevant to astrophysical systems. Similarly to the KHI example, The ablation effects were analyzed by the width of the mixing zone. The morphology of the spikes was affected by the ablation as clearly seen from the experimental images and simulations. A detailed analysis of the structure of the spikes can provide further insights on the ablation effects. In each of the examples above, the CVDL technique would provide a meaningful image similarity order between the sub-figures within the simulation images to the template input of the experimental image. In addition, this technique provides a more extensive and more reliable matching survey, which can decrease the uncertainty margins for the physical parameters, comparing to analysis of the full images only. Therefore, it can serve as a convenient method for analyzing the physical effects and their significance.

The fourth CVDL technique provides a temporal interpolation/extrapolation of experimental results. First, it can provide data augmentation for low-data experiments. Moreover, it can be useful especially when there is a difference between the outputs of the simulations and the experimental images (for example, when the simulation covers a part of the experimental platform) and a prediction of experimental results at additional times are needed. This technique can also be used for an assurance of the model: When there is a series of experimental images, the training can be performed on a part of them. When the training process is accomplished, the prediction of the model for the times of the images that were not selected for the training can be compared for the experimental images from that times. The similarity score between the predicted images and the experimental images can be used for the validity of the model.

A useful feature of the techniques above is that they can be applied on existing simulations databases, since physicists usually perform parameter surveys when analyzing experimental data using simulations. In other words, it can be utilized on previous works without running any additional simulation.

\section{\textit{RayleAI} -- Database Characteristics}
In order to implement the CVDL techniques described above, we first present a state-of-the-art annotated database named \textit{RayleAI} \cite{RayleAI}, which contains thresholded images from a simulation (using the DAFNA hydrodynamic code~\cite{klein2010construction}) of a sinusoidal single-mode RTI perturbation, given by $y=h \mathrm{cos}\left(\frac{2\pi x}{\lambda}\right)$, with a resolution of 64x128 cells, 2.7cm in $x$ axis and 5.4cm in $y$ axis, while each fluid follows the equation of state of an ideal gas (with adiabatic index $\gamma=\frac{5}{3}$) and a hydrostatic equilibrium was set adiabatically with a pressure of 1 bar on the interface. The simulation input consists of three free parameters: Atwood number, gravitational acceleration and the amplitude of the perturbation (as well as additional time parameter). The database contains 101,250 images produced by 1350 different simulations (75 time steps each) with unique set of the free parameters per each simulation. The format of the repository is built upon directories, each represents a simulation performance with the directory name indicating the parameters of the specific simulation. For example, the directory \textit{gravity\_750\_amplitude\_0.5\_atwood\_0.16} is a simulation with $g=750 \frac{\mathrm{cm}}{\mathrm{s}^2}$, initial amplitude of 0.5, and $\A=0.16$. The ranges of the Atwood number, gravity and initial perturbation are presented in Table~\ref{TABLE2}.
Table~\ref{TABLE1} shows the simulation images from DAFNA compared to the experimental images. The simulation images presented are with $g=750 \frac{\mathrm{cm}}{\mathrm{s}^2}$ and $\A=0.16$. 

The choice of these exact parameters was derived from well known experimental results~\cite{waddell2001experimental}. The physical parameters in the experiment were Atwood of $\A=0.155$, gravity of $g=740 \frac{\mathrm{cm}}{\mathrm{s}^2}$ and the initial perturbation wavelength of $0.54$cm. The initial amplitude is not given but can be estimated from the first image by about $0.1$cm. Thus, one can deduce that the simulation in the database with $\A=0.16$, $g=750 \frac{\mathrm{cm}}{\mathrm{s}^2}$ and $h=0.1$ should produce the most accurate result match (as shown in Table \ref{TABLE1}). An example of an experimental image is shown in Fig.~\ref{fig:fullexperiment_2}. The experiment images were originally taken in grayscale. For optimal results, each image was processed with two methods (Erode-Dilate vs. Histogram Equalization), and the most fitting result that resembles the interface between the two fluids best (by expert opinion) was selected, cropped and resized, and then binarized by a threshold. Those images are also included in the database under \textit{/experiment}.


    \begin{minipage}{.45\textwidth}
\begin{table}[H]
    \begin{center}
     \begin{tabular} {  >{\centering\arraybackslash}c |  >{\centering\arraybackslash}c |  >{\centering\arraybackslash}c |  c |  >{\centering\arraybackslash}c }
      \backslashbox{$\textbf{h}$}{$\textbf{T}$} & 0.2 & 0.3 & 0.4\\  
      \hline
        $Exp'$
        &
          \raisebox{-\totalheight}{\includegraphics[height=1.5cm,width=0.75cm]{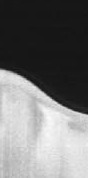}} & 
          \raisebox{-\totalheight}{\includegraphics[height=1.5cm,width=0.75cm]{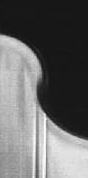}} & 
          \raisebox{-\totalheight}{\includegraphics[height=1.5cm,width=0.75cm]{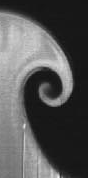}} 
        \\ \hline

        0.1 & 
        \raisebox{-\totalheight}{\includegraphics[height=1.5cm,width=0.75cm]{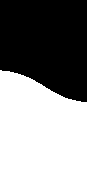}} & 
        \raisebox{-\totalheight}{\includegraphics[height=1.5cm,width=0.75cm]{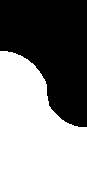}} & 
        \raisebox{-\totalheight}{\includegraphics[height=1.5cm,width=0.75cm]{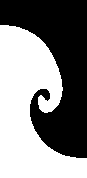}} 
        \\ \hline 

        0.2 &
        \raisebox{-\totalheight}{\includegraphics[height=1.5cm,width=0.75cm]{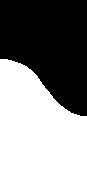}} & 
        \raisebox{-\totalheight}{\includegraphics[height=1.5cm,width=0.75cm]{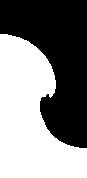}} & 
        \raisebox{-\totalheight}{\includegraphics[height=1.5cm,width=0.75cm]{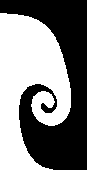}} 
        \\ \hline

      \end{tabular}
      \end{center}
      \vspace{-0.5cm}
      \caption{Diagnostic of RTI in different $T$ and $h$ values from  \textit{RayleAI}.}
      \label{TABLE1}
      \end{table}
    \end{minipage}%
    \hfill
\begin{minipage}{0.45\textwidth}
    \begin{table}[H]
    \centering
  \begin{tabular}
  { | >{\centering\arraybackslash}c |  >{\centering\arraybackslash}c |  >{\centering\arraybackslash}c |  c |  >{\centering\arraybackslash}c |}

      \textbf{Parameter} & \textbf{From} & \textbf{To} & \textbf{Stride} \\  
      \hline
        Atwood ($\A$)
        & 
          0.02 &
          0.5 &
          0.02
        \\ \hline

        Gravity ($g$) [cm/s$^2$] & 
        600 &
        800 &
        25
        \\ \hline 

        Amplitude ($h$) [cm] & 
        0.1 &
        0.5 &
        0.1
        \\ \hline
        X [cm] &
        2.7 &
        2.7 &
        0.0
        \\ \hline
        Y [cm] &
        5.4 &
        5.4 &
        0.0
        \\ \hline
      \end{tabular}
      \caption{Simulation parameters.}
      \vspace{0.39cm}
      \label{TABLE2}
  \end{table}
\vspace{-1.8cm}

\begin{figure}[H]
    \centering
   \includegraphics[width=0.6\linewidth]{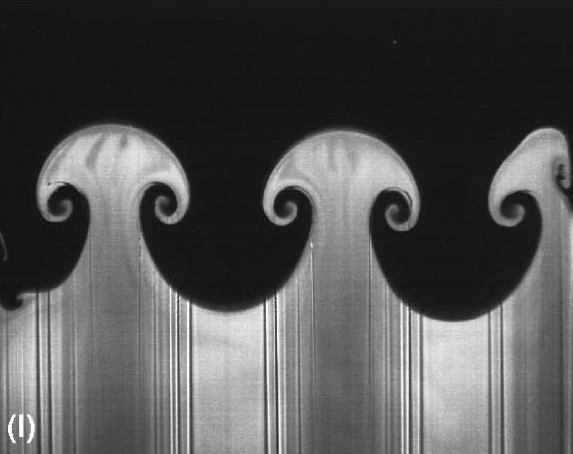}
\caption{The full image from the experiment ($T$=0.4s).}
\label{fig:fullexperiment_2}
\end{figure} 
    \end{minipage}




\section{Deep Computer-Vision Methods} \label{sec:vision_methods}

\subsection{Task \Romannum{1}: Image Retrieval using InfoGAN} \label{sec:LIRE}

Generative Advreserial Networks (GANs) \cite{goodfellow2014generative} is a framework capable to learn a \textit{generator} network $G$, that transforms noise variable $z$ from some noise distribution into a generated sample $G(z)$, while the training of the generator is optimized against a \textit{discriminator} network $D$, which targets to distinguish between real samples with generated ones. The fruitful competition of both $G$ and $D$, in the form of MinMax game, allows $G$ to generate samples such that $D$ will have difficulty with distinguishing real samples between them. The ability to generate indistinguishable new data in an unsupervised manner is one example of a machine learning approach that is able to understand an underlying deep, abstract and generative representation of the data. Information Maximizing Generative Adversarial Network (InfoGAN) \cite{chen2016infogan} utilizes latent code variables $c_i$, which are added to the noise variable. These noise variables are randomly generated from a user-specified domain.
The latent variables impose an Information Theory Regularization term to the optimization problem, which forces $G$ to preserve the information stored in $c_i$ through the generation process. This allows learning interpretative and meaningful representations of the data, with a negligible computation cost, on top of a GAN. The high-abstract-level representation can be extracted from the discriminator (e.g. the last layer before the classification) into a features vector. We use these features in order to measure the similarity between some input image to any other image, by applying some distance function (e.g. $l_2$ norm) on the features of the input to the features of the other image. This methodology provides the ability to order images similarity to a given input image \cite{ganimagesimilarity}.

In order to evaluate InfoGAN performances over \textit{RayleAI}, we also use -- for comparison -- the Computer-Vision technique of LIRE~\cite{lirelibrary}. LIRE is a library that provides image retrieval from databases based on image characteristics among other classic features. LIRE creates a \textit{Lucene} index of image features using both local and global methods. For the evaluation of the similarity of two images, one can calculate their distance in the space they were indexed to. Many state-of-the-art methods for extracting features can be used, such as Gabor Texture Features \cite{zhang2000content}, Tamura Features \cite{tamurafeatures}, or FCTH \cite{chatzichristofis2008fcth}. For our purposes, we found that Tamura Features method is the most suitable method that LIRE provides, as it indexes \textit{RayleAI} images in a more dispersed fashion. The Tamura feature vector of an image is an 18 double values descriptor that represents texture features in the image that correspond to human visual perception.

\subsection{Task \Romannum{2}: Parameters Regression using ConvNet -- pReg} \label{pReg:explain}
Many Deep Learning techniques obtain state-of-the-art results for regression tasks, in a wide range of CV applications \cite{lathuiliere2019comprehensive} such as Pose Estimation, Facial Landmark Detection, Age Estimation, Image Registration and Image Orientation \cite{fischer2015image} \cite{mahendran20173d}. Most of the deep learning architectures used for regression tasks on images are Convolutional Neural Networks (ConvNets), which are usually composed of blocks of Convolutional layers followed by a Pooling layer, and finally Fully-Connected layers. The dimension of the output layer depends on the task, and its activation function is usually linear or sigmoid. 

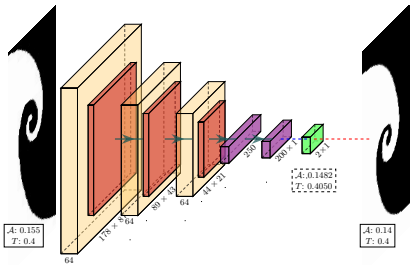
\begin{figure}[H]
    \centering
\resizebox{0.45\textwidth}{!}{
\begin{tikzpicture}
\tikzstyle{connection}=[ultra thick,every node/.style={sloped,allow upside down},draw=\edgecolor,opacity=0.7]
\tikzstyle{copyconnection}=[ultra thick,every node/.style={sloped,allow upside down},draw={rgb:blue,4;red,1;green,1;black,3},opacity=0.7]

\node[canvas is zy plane at x=0] (temp) at (-2.2,0,0) {\includegraphics[width=4.5cm,height=8cm]{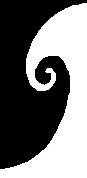}};

\pic[shift={(0,0,0)}] at (0,0,0) 
    {Box={
        name=conv1,
        caption= ,
        xlabel={{64, }},
        zlabel=$178\times87$,
        fill=\ConvColor,
        height=30,
        width=3,
        depth=30
        }
    };

\pic[shift={ (0,0,0) }] at (conv1-east) 
    {Box={
        name=pool1,
        caption= ,
        fill=\PoolColor,
        opacity=0.5,
        height=20,
        width=1,
        depth=20
        }
    };

\pic[shift={(1,0,0)}] at (pool1-east) 
    {Box={
        name=conv2,
        caption= ,
        xlabel={{64, }},
        zlabel=$89\times43$,
        fill=\ConvColor,
        height=20,
        width=3,
        depth=20
        }
    };

\draw [connection]  (pool1-east)    -- node {\midarrow} (conv2-west);

\pic[shift={ (0,0,0) }] at (conv2-east) 
    {Box={
        name=pool2,
        caption= ,
        fill=\PoolColor,
        opacity=0.5,
        height=15,
        width=1,
        depth=15
        }
    };

\pic[shift={(1,0,0)}] at (pool2-east) 
    {Box={
        name=conv3,
        caption= ,
        xlabel={{64, }},
        zlabel=$44\times21$,
        fill=\ConvColor,
        height=15,
        width=3,
        depth=15
        }
    };

\draw [connection]  (pool2-east)    -- node {\midarrow} (conv3-west);

\pic[shift={ (0,0,0) }] at (conv3-east) 
    {Box={
        name=pool3,
        caption= ,
        fill=\PoolColor,
        opacity=0.5,
        height=10,
        width=1,
        depth=10
        }
    };

\pic[shift={(1,0,0)}] at (conv3-east) 
    {Box={
        name=soft1,
        caption=,
        xlabel={{" ","dummy"}},
        zlabel=$250\time1$,
        fill=\FcReluColor,
        opacity=0.8,
        height=3,
        width=1.5,
        depth=15
        }
    };

\draw [connection]  (conv3-east)    -- node {\midarrow} (soft1-west);

\pic[shift={(1,0,0)}] at (soft1-east) 
    {Box={
        name=soft2,
        caption=,
        xlabel={{" ","dummy"}},
        zlabel=$200\times 1$,
        fill=\FcReluColor,
        opacity=0.8,
        height=3,
        width=1.5,
        depth=10
        }
    };

\draw [connection]  (soft1-east)    -- node {\midarrow} (soft2-west);

\pic[shift={(1,0,0)}] at (soft2-east) 
    {Box={
        name=out,
        caption=,
        xlabel={{" ","dummy"}},
        zlabel=$2\times 1$,
        fill=\OutColor,
        opacity=0.8,
        height=3,
        width=1.5,
        depth=6
        }
    };

\draw[dashed, ultra thick,every node/.style={sloped,allow upside down},draw=blue,opacity=0.7] (soft2-east) -- (out-west);

\node[draw, align=center, dashed] at (8, -1.5)  (params)  {$\A$: 0.1482\\$T$: 0.4050};

\node[canvas is zy plane at x=0] (temp) at (10.6,0,0) {\includegraphics[width=4.5cm,height=8cm]{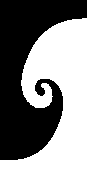}};

\draw[dashed,every node/.style={sloped,allow upside down},draw=red] (8,0,0) --  (10,0,0);

\node[draw, align=center] at (-2.5, -3.5)  (o1)  {$\A$: 0.155\\$T$: 0.4};

\node[draw, align=center] at (10.3, -3.5)  (o2)  {$\A$: 0.14\\$T$: 0.4};

\end{tikzpicture}
}

    \caption{CNN model, named \textit{pReg}, for $\A$ and $T$ parameters Regression. The left image is an example of an experimental input, with the real parameters of $\A$: 0.155, $T$: 0.4, which the model predicts for the values $\A$: 0.1482, $T$: 0.4050 as can be seen under the green layer. The red dotted line indicates the similarity search operation that quantifies the distance of images from \textit{RayleAI}, based on the $l_2$ distance over $\A$ and $T$.}
    \label{fig:regression_arch}

\end{figure}

ConvNets can be used for retrieving the parameters of an experiment image, via regression. Our model (henceforth \textit{pReg}) (Fig. \ref{fig:regression_arch}) consists of 3 Convolutional layers with 64 filters, with a kernel size $5\times 5$, and with $l_2$ regularization, each followed by a Max-Pooling layer, a Dropout of $0.1$ rate and finally Batch Normalization. Then, there are two Fully-Connected layers of 250 and 200 features, which are separated again by a Batch Normalization layer. Finally, the Output layer of our network has 2 features (as will described next), and is activated by sigmoid to prevent the exploding gradients problem. Since the most significant parameters for describing each image frame are Amplitude and Time -- which \textit{pReg} is trained to predict -- we used only a subset of \textit{RayleAI} for the training set, namely images with the following parameters: $\A$ $\in [0.08, 0.5]$ (with a stride of $0.02$), $g$ $\in \{625, 700, 750, 800\}$, $h$ $\in [0.1, 0.5]$ (with a stride of $0.1$) and $T$ $\in [0.1, 0.6]$ (with a stride of $0.01$). We fixed a small amount of values for Gravity and for Amplitude, so the network will not try to learn the variance that these parameters impose while expanding our database with as minimal noise as possible. We chose the value ranges of Atwood and Time in order to expose the model to images with both small and big perturbations, such that the amount of the latter ones will not be negligible. Our reduced training set consists of $\sim 16K$ images, and our validation set consists of $\sim 4K$ images. Nonetheless, for increasing generalization and for decreasing model overfitting, we employed data augmentation. Since there is high significance for the perspective from which each image is taken, the methods of data augmentation should be carefully chosen: Rotation, shifting and flipping methods may generate images such that the labels of the original parameters do not fit for them. Therefore, we augment our training set with only zooming in/out (zoom range=0.1) via TensorFlow \cite{abadi2016tensorflow} preprocessing.

\subsection{Task \Romannum{3}: Quality-Aware Template Matching using QATM}
\label{QATM:explain}
One variation of the Template Matching problem is defined as follows: Given an exemplar image $E$, find the most similar region of interest in a target image $S$ \cite{brunelli2009template}. Classic template matching methods often use Sum-of-Squared Differences (SSD) or Normalized Cross-Correlation (NCC) to asses the similarity score between a template and an underlying image. These approaches work well when the transformation between the template and the target search image is simple. However, with non-rigid transformations, which are common in real-life, they start to fail. Quality-Aware Template Matching (QATM) \cite{qatm} method is a standalone template matching algorithm and a trainable layer with trainable parameters that can be embedded into any Deep Neural Network. QATM is inspired by assessing the matching quality of the source and target templates. It defines the $QATM(e,s)$-measure as the product of likelihoods that a patch $s$ in $S$ is matched in $E$ and a patch $e$ in $E$ is matched in $S$. Once $QATM(e, s)$ is computed, we can compute the template matching map for the template image $E$ and the target searched image $S$. Eventually, we can find the best-matched region ${R^*}$ which maximizes the overall matching quality. Therefore, the technique is of great need when templates are complicated and targets are noisy. Thus most suitable for RTI images from simulations and experiments.

\subsection{Task \Romannum{4}: Time Series Prediction using PredRNN} \label{sec:PredRNNexplain}
Learning the evolution of the RTI in order to predict future time or gap frames requires both understanding of the spatial aspects of each time frame (e.g. the interface between the fluids), and understanding of time development: As the time progresses, the simulation tends to be more and more chaotic. Convolutional Long Short Term Memory networks (CLSTMs)~\cite{xingjian2015convolutional} is a class of algorithms which able to predict future image states by past and present image states based on training sequences of images.
The architecture of this network is based on a two-dimensional grid of units that pass spatial information vertically (upwards), and temporal information horizontally (rightwards).
However, the standard CLSTMs architectures lack the capability of preserving the temporal information for long terms, since the spatial information that is learned via the top unit in a specific time step, is not passed to the bottom unit in the next time step, leading to the loss of important information.
PredRNN~\cite{wang2017predrnn} is a state-of-the-art Recurrent Neural Network for predictive learning using LSTMs. PredRNN memorizes both spatial appearances and temporal variations in a unified memory pool. Unlike standard LSTMs, and in addition to the standard memory transition within them, memory in PredRNN can travel through the whole network in a zigzag direction, therefore from the top unit of some time step to the bottom unit of the other. Thus, PredRNN is able to preserve the temporal as well as the spatial memory for long-term motions. In this work, we use PredRNN for predicting future time steps of simulations as well as experiments, based on the given sequence of time steps.

\section{Evaluation Methodology} \label{sec:phy_loss}

In order to test how the discussed above techniques perform on physical simulations as well as experiments, we propose new task-specific test methods, that quantify how well each technique operates on a concrete database.
We present novel evaluation methodologies for the techniques presented in Sections ~\ref{sec:LIRE}, \ref{pReg:explain} and \ref{QATM:explain}, based on a suitable corresponding loss measure for the first two tasks, and a sophisticated clustering-visualization method for the third. The evaluation of last forth technique (\ref{sec:PredRNNexplain}) will be discussed separately. 

The first evaluation method, namely \textit{Physical Loss}, quantifies \textit{how meaningful the results of the technique are}, i.e. whether the results of the technique are reflected in the (physical) annotations of the data. For example, in the case of Image Retrieval, it is inconclusive to decide whether the results are sufficient solely based on visual examination, since it is a very difficult task for humans to determine the correct ordering of lots of results, thus it is hard to establish whether the technique is satisfactory. Therefore, we suggest to measure how each input image is \textit{physically close} to each of the returned image outputs, based on some or all of their parameters labels. Thus, for Image Retrieval (and for Parameters Regression, explained later in Section \ref{pReg:results_sec}), each output image gets two scores -- one from the technique at hand, e.g. similarity score, and one from the difference between its parameters to the parameters of the input image. In cases of high correlations of these scores, one can infer that indeed the results of the technique are of a meaningful (physical) value. We chose the Mean Squared Error (MSE) as the parameter difference function, although it may be calculated via any desired error function. We note that since the ranges of the parameters are scaled differently, we suggest to normalize them beforehand.

Furthermore, one important aspect that results from \textit{Physical Loss} is the ability to identify the parameters which are likely to produce a small impact on the simulation results (depending on time). For example, in the case of small ratio between the amplitude and the wavelength of the perturbation (up to a few percent), RTI grows linearly according to Eq.~\ref{eq:linear} and approximately preserves its initial shape. Therefore, two simulations that differ only by their initial small amplitudes will practically result in the same late evolution up to a constant time shift. As a result, it is expected from physical considerations that if one produces an amplitude-based \textit{Physical Loss} methodology for late times, the CVDL techniques will generate semi-random values of error as the amplitude hardly affect the simulation in late times. A similar result is also expected for the gravity parameter since for incompressible fluids (a good approximation in our case), two simulations that differ only by the gravity parameter will practically result in the same evolution as a function of the normalized time. A useful definition of the normalized time is $\tilde{t}=\sqrt{\frac{A g}{\lambda}} t$ as also reflected from Eq.~\ref{eq:linear}. Concluding the above physical influence of the initial amplitude and the gravity parameters, only the Atwood and time parameters should have a significant impact on the results and are expected to be identified using the physical loss methodology. 

In cases where there are no meaningful (physical) annotations, we developed another new evaluation method. Specifically in the case of Template Matching, where some partial template is searched through a database. Unlike the physical loss case, the physical parameters of the returned partial region of interest have no unique physical labels, since we might expect to find this template in images from a wide range of different parameters. For that end, we present a relaxed-evaluation method, that quantifies how well the technique at hand separates similar images from dissimilar images. Similarly to our \textit{Physical Loss} methodology, we use two values for each output: The score from the technique, and a cluster number -- returned from some unsupervised clustering algorithm. Scenarios in which continuous sequences from the results of the technique are from the same cluster might indicate the ability of the technique to perform a proper distinction between classes of similarity to a given input template. Alternatively, cases of sequences of results from mixed clusters, especially in the first and most similar regions, might prove that the technique did not succeed in separating the most similar images from the rest.
Accordingly, we applied K-Means as our clustering algorithm, after extracting the main features from each image through Principal Component Analysis (PCA) to achieve more precise results \cite{ding2004k}. Next, we present in Section \ref{sec:results} the evaluation results of the techniques from Section \ref{sec:vision_methods}.

\section{Results and Discussion} \label{sec:results}
\subsection{Task \Romannum{1}: Image Retrieval}
In order to test the performance of InfoGAN against LIRE we chose two separate test cases. In the first general test case, we chose 13,000 random input test images from \textit{RayleAI}. In the latter, we chose approximately 10,000 input images with $T>0.25$ in order to pick the most \textit{complex} images, as the RTI is more chaotic and dominant in this regime. We then executed InfoGAN and LIRE on the entire \textit{RayleAI} data-set for both test cases. Then, for each tested image and for each tool, we sorted the results according to the similarity scores that were given by the model. To evaluate the results and quantify how well the tool performed, we employed the \textit{physical loss} methodology, introduced in section~\ref{sec:phy_loss}, over the Atwood parameter. Then, for each tool and test case, we calculated the average physical loss. 

\begin{figure}[H]
\centering
\begin{subfigure}[b]{.24\linewidth}
\centering
\includegraphics[width=1.1\linewidth]{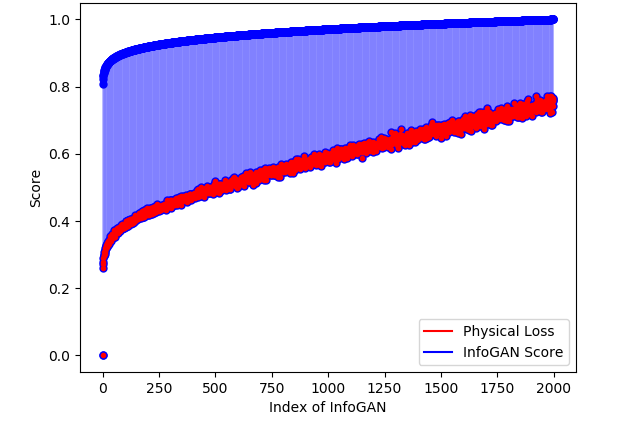}
\caption{\centering \scalebox{0.8}{Comp' InfoGAN}} \label{InfoGan:longtimes}
\end{subfigure}
\begin{subfigure}[b]{.24\linewidth}
\centering
\includegraphics[width=1.1\linewidth]{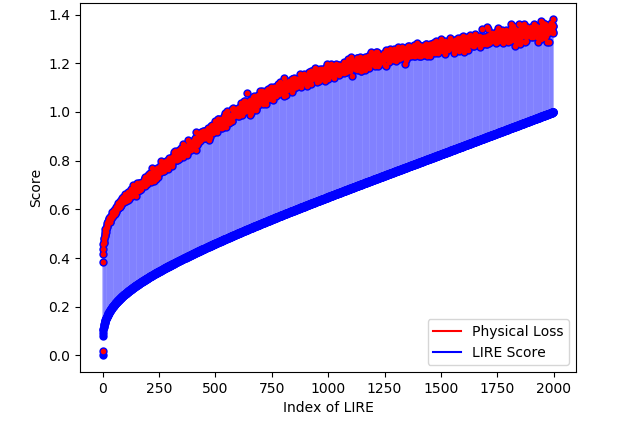}
\caption{\centering \scalebox{0.8}{Comp' LIRE}} \label{Lire:longtimes}
\end{subfigure}
\begin{subfigure}[b]{.24\linewidth}
\centering
\includegraphics[width=1.1\linewidth]{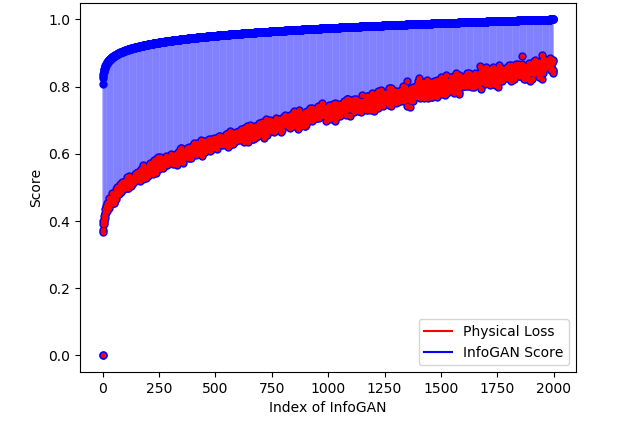}
\caption{\centering \scalebox{0.8}{Rand' InfoGAN}} \label{InfoGan:random}
\end{subfigure}
\begin{subfigure}[b]{.24\linewidth}
\centering
\includegraphics[width=1.1\linewidth]{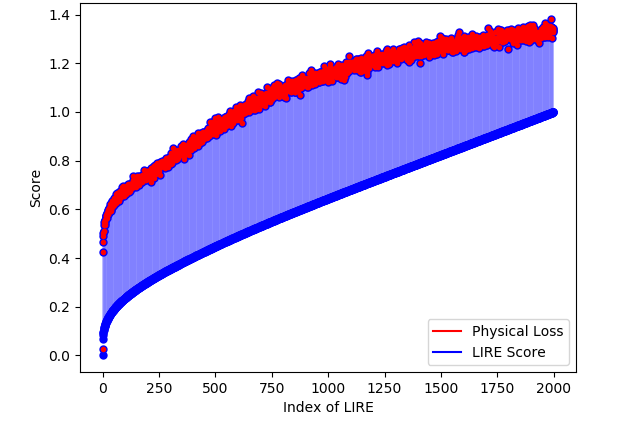}
\caption{\centering \scalebox{0.8}{Rand' LIRE}} \label{Lire:random}
\end{subfigure}
\caption{LIRE and InfoGAN averaged \textit{physical loss} methodology over Atwood.}
\label{all_plots_tests}
\end{figure}

In Fig. \ref{all_plots_tests} we present the physical loss methodology (using a comparison between the technique score [in blue] and the physical loss [in red] per each index, and draw a thin blue line to correlate them) only on the Atwood parameter, as it is the most significant parameter. 
In Figures \ref{InfoGan:longtimes} and \ref{Lire:longtimes}, we observe that InfoGAN outperformed LIRE on the complex images test, as the averaged physical loss of the first -- and most important -- indices of InfoGAN is $\sim 0.25$, in contrast to $\sim 0.4$ of LIRE. Furthermore, InfoGAN outperforms LIRE along the entire 2000 first examined indices, showing many powerful capabilities in the complex data case.
In Figures \ref{InfoGan:random} and \ref{Lire:random}, we can see that InfoGAN and LIRE perform quite the same in the first indices, with averaged physical loss of around 0.4. Yet, if we focus on the entire 2000 first indices, we see that InfoGAN starts to outperform LIRE with smaller physical loss values. 
Additionally, it seems that there is a higher correlation between the scores of InfoGAN to their corresponding physical loss values (blue and red lines act accordingly) in each of the test cases, which indicated again on the ability of InfoGAN to learn the underlying physical pattern of the data. Another important aspect in which InfoGAN outperforms LIRE in both test-cases is the width of the physical loss line: As the red line is thinner, there is less noise and therefore the results have more physical sense. 
Although it seems that all red lines are of approximately the same width, the lines of InfoGAN are much thinner since the score ranges of InfoGAN are smaller than the ones of LIRE (scaled from 0 to 1, in contrast to LIRE which are scaled from 0 to 1.4).
Note, that in all Figures, the blue lines are normalized by the min-max normalization method, contrary to the red lines which are presented as raw values.
The overwhelming superiority of InfoGAN is somehow expected and can be explained as the ability of a deep learning model to learn complex patterns from our tailor-made and trained database. However, although LIRE provides decent results without requiring to be trained on a specific organized database (which obtaining is not always an easy task), it is still a classic image processing tool, which lacks the learning capabilities that will show it to understand deep patterns from the data (examplification in Table \ref{image_similarity_table}). Therefore, for image retrieval applications with suitable databases, we suggest applying InfoGAN.

\begin{table}[H]
    \begin{center}
    \begin{tabular}{c|c|c|c|c|c|c|c|c|c|c|c|c}
        {$\textbf{Index}$} & 1 & 2 & 3 & 4 & 5 & 6 & 7 & 8 & 9 & 10 & 11 & 12\\  
        \hline
    $\begin{aligned}\\\text{LIRE}\\ \includegraphics[height=1.5cm,width=0.75cm]{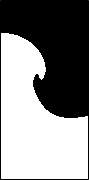}\\\text{InfoGAN}\\\end{aligned}$ &
    
    $\underset{\includegraphics[height=1.5cm,width=0.75cm]{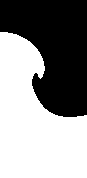}}{\includegraphics[height=1.5cm,width=0.75cm]{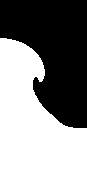}}$ &
    $\underset{\includegraphics[height=1.5cm,width=0.75cm]{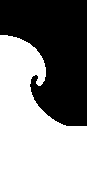}}{\includegraphics[height=1.5cm,width=0.75cm]{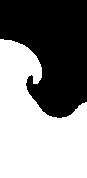}}$&
    $\underset{\includegraphics[height=1.5cm,width=0.75cm]{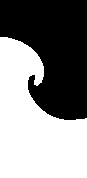}}{\includegraphics[height=1.5cm,width=0.75cm]{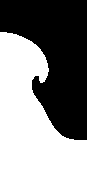}}$ &
    $\underset{\includegraphics[height=1.5cm,width=0.75cm]{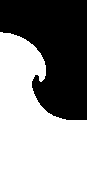}}{\includegraphics[height=1.5cm,width=0.75cm]{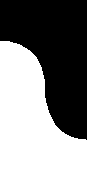}}$ &
    $\underset{\includegraphics[height=1.5cm,width=0.75cm]{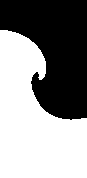}}{\includegraphics[height=1.5cm,width=0.75cm]{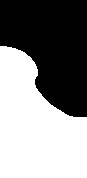}}$ &
    $\underset{\includegraphics[height=1.5cm,width=0.75cm]{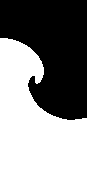}}{\includegraphics[height=1.5cm,width=0.75cm]{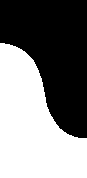}}$ &
    $\underset{\includegraphics[height=1.5cm,width=0.75cm]{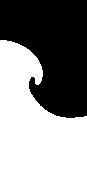}}{\includegraphics[height=1.5cm,width=0.75cm]{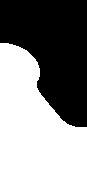}}$&
    $\underset{\includegraphics[height=1.5cm,width=0.75cm]{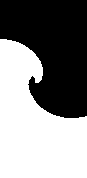}}{\includegraphics[height=1.5cm,width=0.75cm]{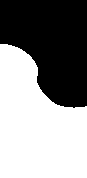}}$ &
    $\underset{\includegraphics[height=1.5cm,width=0.75cm]{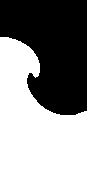}}{\includegraphics[height=1.5cm,width=0.75cm]{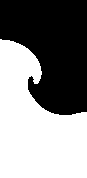}}$ &
    $\underset{\includegraphics[height=1.5cm,width=0.75cm]{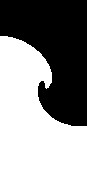}}{\includegraphics[height=1.5cm,width=0.75cm]{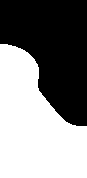}}$ &
    $\underset{\includegraphics[height=1.5cm,width=0.75cm]{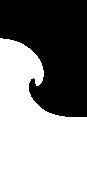}}{\includegraphics[height=1.5cm,width=0.75cm]{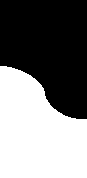}}$ &
    $\underset{\includegraphics[height=1.5cm,width=0.75cm]{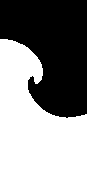}}{\includegraphics[height=1.5cm,width=0.75cm]{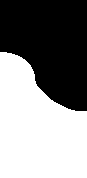}}$
    \\\hline
      $\begin{aligned}[m] \\\text{LIRE}\\ \begin{aligned}[m]\includegraphics[height=1.5cm,width=0.75cm]{{Image_Similarity_Table/test_image_2/test_img_2}.png} \end{aligned} \\\text{InfoGAN}\\\end{aligned} $  &
  
    $\underset{\includegraphics[height=1.5cm,width=0.75cm]{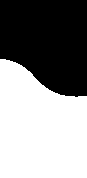}}{\includegraphics[height=1.5cm,width=0.75cm]{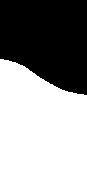}}$ &
    $\underset{\includegraphics[height=1.5cm,width=0.75cm]{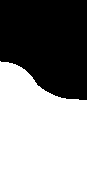}}{\includegraphics[height=1.5cm,width=0.75cm]{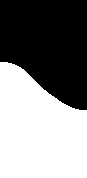}}$&
    $\underset{\includegraphics[height=1.5cm,width=0.75cm]{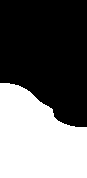}}{\includegraphics[height=1.5cm,width=0.75cm]{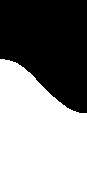}}$ &
    $\underset{\includegraphics[height=1.5cm,width=0.75cm]{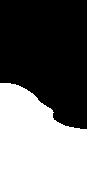}}{\includegraphics[height=1.5cm,width=0.75cm]{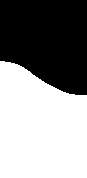}}$ &
    $\underset{\includegraphics[height=1.5cm,width=0.75cm]{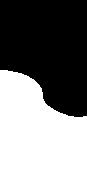}}{\includegraphics[height=1.5cm,width=0.75cm]{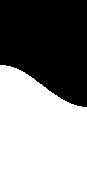}}$ &
    $\underset{\includegraphics[height=1.5cm,width=0.75cm]{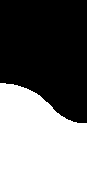}}{\includegraphics[height=1.5cm,width=0.75cm]{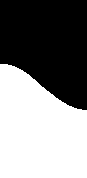}}$ &
    $\underset{\includegraphics[height=1.5cm,width=0.75cm]{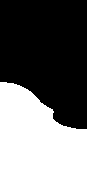}}{\includegraphics[height=1.5cm,width=0.75cm]{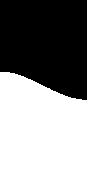}}$&
    $\underset{\includegraphics[height=1.5cm,width=0.75cm]{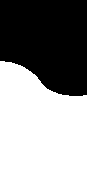}}{\includegraphics[height=1.5cm,width=0.75cm]{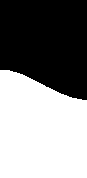}}$ &
    $\underset{\includegraphics[height=1.5cm,width=0.75cm]{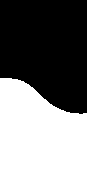}}{\includegraphics[height=1.5cm,width=0.75cm]{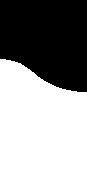}}$ &
    $\underset{\includegraphics[height=1.5cm,width=0.75cm]{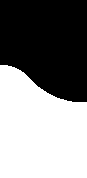}}{\includegraphics[height=1.5cm,width=0.75cm]{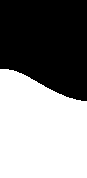}}$ &
    $\underset{\includegraphics[height=1.5cm,width=0.75cm]{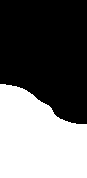}}{\includegraphics[height=1.5cm,width=0.75cm]{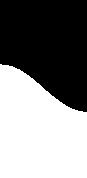}}$ &
    $\underset{\includegraphics[height=1.5cm,width=0.75cm]{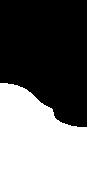}}{\includegraphics[height=1.5cm,width=0.75cm]{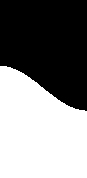}}$
    \\\hline 
    \end{tabular}
    \end{center}
    \vspace{-0.5cm}
    \caption{LIRE and InfoGAN first 12 results for two chosen test images.}
    \label{image_similarity_table}
    \end{table}

\subsection{Task \Romannum{2}: Parameters Regression} \label{pReg:results_sec}
In order to test the performances of our \textit{pReg} network, we employed evaluation tests that are similar to the tests presented in section \ref{sec:phy_loss}. We used \textit{pReg} to predict the -- activated by sigmoid -- parameters: $\A$ and $T$ for $2,000$ random images. Then, for each image we searched through \textit{RayleAI} for the $2,000$ images with the lowest scores, based on their $l_2$ distance between their $\A$ and $T$ activated by sigmoid parameters, to that of the input image.

\vspace{-1cm}

\begin{figure}[H]
\centering
\begin{subfigure}[b]{.31\linewidth}
\centering
\includegraphics[width=1.0\linewidth]{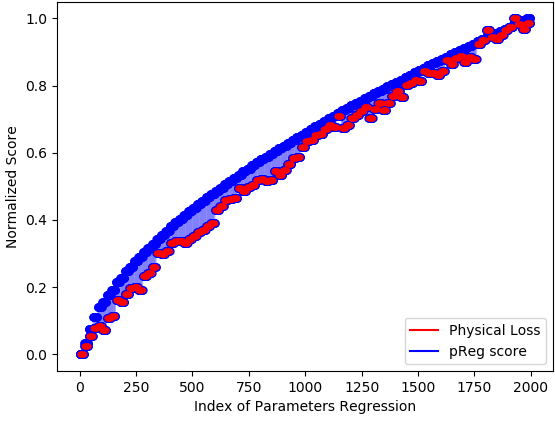}
\caption{\scalebox{0.8}{Averaged $T$}} \label{pRegOut:time}
\end{subfigure}
\hspace{0.08cm}
\begin{subfigure}[b]{.31\linewidth}
\centering
\includegraphics[width=1.0\linewidth]{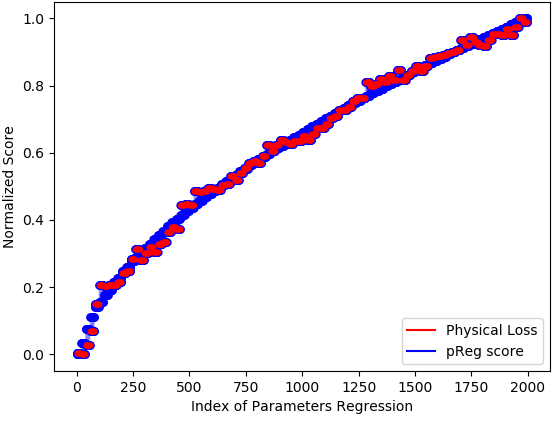}
\caption{\scalebox{0.8}{Averaged $\A$}} \label{pRegOut:at}
\end{subfigure}
\hspace{0.08cm}
\begin{subfigure}[b]{.31\linewidth}
\centering
\includegraphics[width=1.0\linewidth]{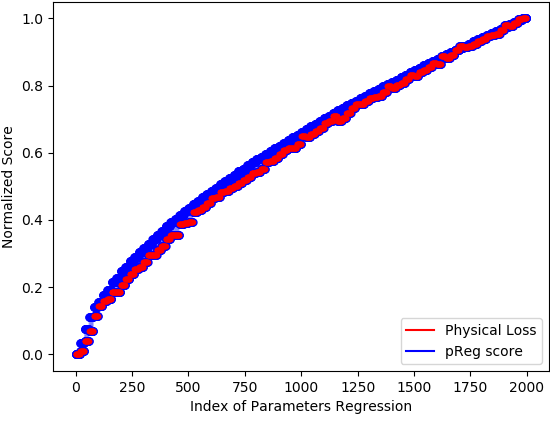}
\caption{\scalebox{0.8}{Averaged $\A$ \& $T$}} \label{pRegOut:time_at}
\end{subfigure}

\caption{\textit{pReg} averaged \textit{physical loss} methodology over Atwood and time.} \label{pRegOut:all_plots}
\end{figure}
\begin{table}[ht]
    \begin{center}
     \begin{tabular} { | c | c | c |  c | }
     
    \hline 
     \begin{tabular} {l} $g$:-800,$h$:0.5, \\ $\A$:0.34,$T$:0.37 \end{tabular} & 
     \begin{tabular} {l} $g$:-625,$h$:0.5, \\ $\A$:0.34,$T$:0.36 \end{tabular} & 
     \begin{tabular} {l} $g$:-800,$h$:0.2, \\ $\A$:0.34,$T$:0.36 \end{tabular} & 
     \begin{tabular} {l} $g$:-800,$h$:0.3, \\ $\A$:0.34,$T$:0.36 \end{tabular}
     \\  
        
      \hdashline
          \includegraphics[height=1.5cm,width=0.75cm]{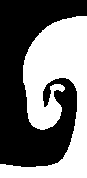} &
          \includegraphics[height=1.5cm,width=0.75cm]{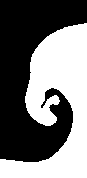} &
          \includegraphics[height=1.5cm,width=0.75cm]{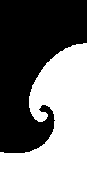} &
          \includegraphics[height=1.5cm,width=0.75cm]{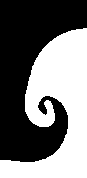}
          
          \\ 
          \hdashline
          input & 1 & 2 & 3
          
        \\ \hline \hline
        \begin{tabular} {l} $g$:-700,$h$:0.3, \\ $\A$:0.34,$T$:0.36 \end{tabular} & 
        \begin{tabular} {l} $g$:-625,$h$:0.1, \\ $\A$:0.34,$T$:0.36 \end{tabular} &
        \begin{tabular} {l} $g$:-750,$h$:0.5, \\ $\A$:0.34,$T$:0.36 \end{tabular} & 
        \begin{tabular} {l} $g$:-750,$h$:0.2, \\ $\A$:0.34,$T$:0.36 \end{tabular} 
        \\  
        
      \hdashline

        \includegraphics[height=1.5cm,width=0.75cm]{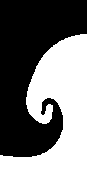} &
        \includegraphics[height=1.5cm,width=0.75cm]{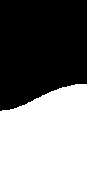} &
         \includegraphics[height=1.5cm,width=0.75cm]{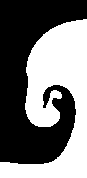} &
         \includegraphics[height=1.5cm,width=0.75cm]{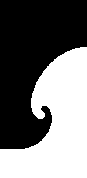}
          \\ 
          \hdashline
          4 & 5 & 6 & 7
          
        \\ \hline \hline
        \begin{tabular} {l} $g$:-750,$h$:0.1, \\ $\A$:0.34,$T$:0.36 \end{tabular} & 
        \begin{tabular} {l} $g$:-625,$h$:0.3, \\ $\A$:0.34,$T$:0.36 \end{tabular} & 
        \begin{tabular} {l} $g$:-800,$h$:0.4, \\ $\A$:0.34,$T$:0.36 \end{tabular} & 
        \begin{tabular} {l} $g$:-750,$h$:0.3, \\ $\A$:0.34,$T$:0.36 \end{tabular} 
        \\ 
        \hdashline

        \includegraphics[height=1.5cm,width=0.75cm]{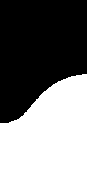} &
         \includegraphics[height=1.5cm,width=0.75cm]{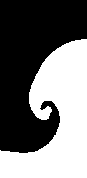} &
         \includegraphics[height=1.5cm,width=0.75cm]{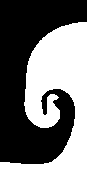} &
         \includegraphics[height=1.5cm,width=0.75cm]{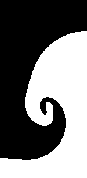}

          \\ 
          \hdashline
          8 & 9 & 10 & 11
          
          \\ \hline \hline
        \begin{tabular} {l} $g$:-700,$h$:0.5, \\ $\A$:0.34,$T$:0.36 \end{tabular} & 
        \begin{tabular} {l} $g$:-700,$h$:0.2, \\ $\A$:0.34,$T$:0.36 \end{tabular} &
        \begin{tabular} {l} $g$:-750,$h$:0.4, \\ $\A$:0.34,$T$:0.36 \end{tabular} & 
        \begin{tabular} {l} $g$:-625,$h$:0.4, \\ $\A$:0.34,$T$:0.36 \end{tabular} 
        \\  
        
      \hdashline

        \includegraphics[height=1.5cm,width=0.75cm]{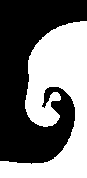} &
        \includegraphics[height=1.5cm,width=0.75cm]{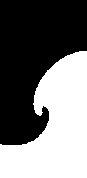} &
        \includegraphics[height=1.5cm,width=0.75cm]{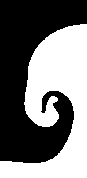} &
        \includegraphics[height=1.5cm,width=0.75cm]{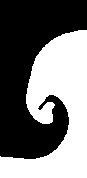}
          \\ 
          \hdashline
          12 & 13 & 14 & 15 
          
        \\ \hline
      \end{tabular}
      \end{center}
      \vspace{-0.5cm}
      \caption{pReg Similarity based on parameters regression.}
      \label{pReg_table}
      \end{table}

As can be seen in Fig. \ref{pRegOut:time}, the model explains very well the time, especially in the lowest ($\leq 500$) and highest ($\geq 1800$) indices, since the red dots of the normalized \textit{physical loss} over the time, and the blue dots of the normalized $l_2$ distance from the predicted parameters, act similarly. The higher difference between the dots in the middle of the scale ($500 <$ indices $< 1800$) is somehow expected, as it is harder for models to predict the parameters accurately in cases where there is a 'mild' physical difference. Yet, in cases where there is a high resemblance or significant difference with respect to the physical loss, it is more likely that the model will predict similar parameters or very different parameters, respectively. In Fig. \ref{pRegOut:at}, we can see that the model explains even better the Atwood parameter, as the graphs are almost the same with some small noise. This can be explained by the significance and importance of the Atwood parameter. In Fig. \ref{pRegOut:time_at} we see that the combination of Atwood and time greatly outperforms the former two cases, since the red line almost converges to the blue line.
We note that as the predicted parameters in \textit{pReg} are only $\A$ and $T$, and the difference is calculated only over them, for each input image there are lots of images in \textit{RayleAI} with the same calculated distance -- same $\A$ and $T$, but different $g$ or $h$. Therefore the trends in Figs. \ref{pRegOut:at}, \ref{pRegOut:time}, \ref{pRegOut:time_at} might have dense blocks of dots -- with same or very similar scores. Furthermore, the fact that there are lots of images with the same loss results in highly similar median and average values trends, therefore we present only the average values graphs from space considerations. Finally, since images with the same $\A$ and $T$ but different $g$ or $h$ have the same loss (over $\A$ and $T$), they are ordered arbitrarily. Therefore, the physical loss over all parameters does not explain the similarity order of \textit{pReg}, because of the arbitrariness that $g$ and $h$ impose. In Table \ref{pReg_table} we present an example for image similarity search based on physical parameters regression.

\subsection{Task \Romannum{3}: Template Matching}
For the evaluation of the QATM algorithm we cropped 16 templates from the experiment images. For each template, we employed the following procedure: We ran the QATM algorithm (1-to-1 version, the most precised one) on each image in \textit{RayleAI} and found a matched sub-figure. Then, we sorted the results in accordance to the QATM similarity scores. For results evaluation, we employed the loss methodology of PCA and K-Means, as described in Section \ref{sec:phy_loss}.

\begin{figure}[H]
\centering
\begin{subfigure}[b]{.3\linewidth}
\centering
\includegraphics[width=1.1\linewidth]{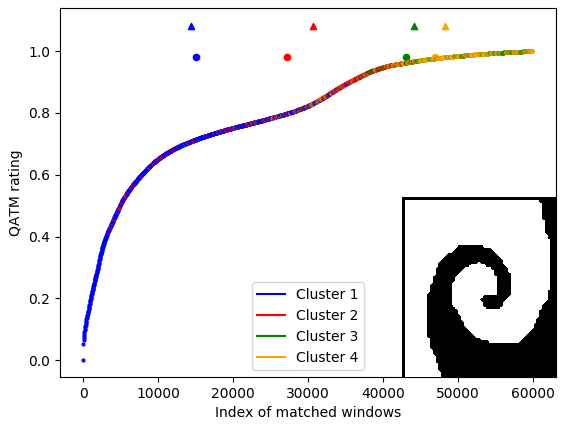}
\caption{\scalebox{0.8}{Unique Template}} \label{QATM:good}
\end{subfigure}
\hspace{0.08cm}
\begin{subfigure}[b]{.3\linewidth}
\centering
\includegraphics[width=1.1\linewidth]{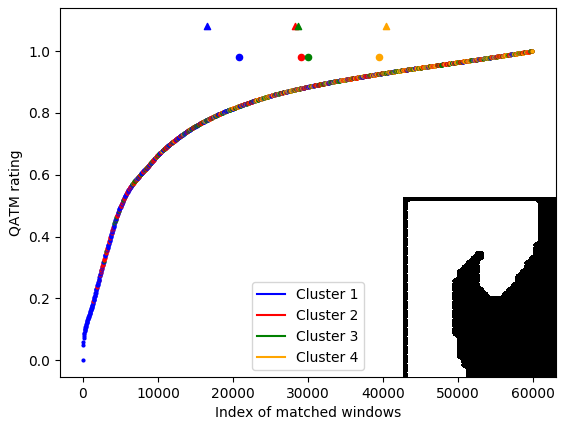}
\caption{\scalebox{0.8}{Semi-Unique Template}} \label{QATM:middle}
\end{subfigure}
\hspace{0.08cm}
\begin{subfigure}[b]{.3\linewidth}
\centering
\includegraphics[width=1.1\linewidth]{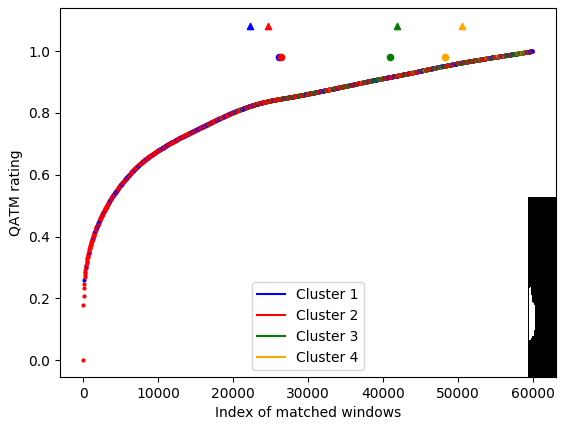}
\caption{\scalebox{0.8}{Non-Unique Template}} \label{QATM:bad}
\end{subfigure}
\caption{PCA and k-means clustering methodology made on QATM results.} \label{QATMt:all_plots}
\end{figure}

In Fig. \ref{QATMt:all_plots} we present the results of three representative templates, while in each the normalized score results of the algorithm are sorted in an increasing order. The color of every point represents the cluster of the underlying sub-figure, returned by the K-Means algorithm. The up-arrows and the circles lying above the curves represent the median and the average of the indices of each cluster, respectively. To keep the trends readable, only one of each 30 dots is presented. As can be seen in Fig. \ref{QATM:good}, the clustering algorithm divides the indices into 4 separate and distinct areas: There is pure congestion of blue dots in the first thousands of indices, without any rogue non-blue dots. This indicates that the algorithm understood the template successfully and found lots of significantly similar sub-figures. This powerful result might be explained because the searched template is of a 'unique' shape, which helps QATM extract lots of features and compare them to \textit{RayleAI}.
in Fig. \ref{QATM:middle}, we can see pure congestion of blue dots in the first hundreds of indices, and a mixture of blue and red dots, with the unignorable presence of green dots in the right following indices. This mixture of clusters, that appears in relatively small indices, indicates that the algorithm's results start to be less meaningful after a couple of hundreds of indices. This can be explained since the searched template is less unique than the previous template.
Finally, in Fig. \ref{QATM:bad} the blue and red clusters seem to be inseparable all along the index axis. This indicates that the algorithm did not understand well the template and has difficulties to bring quality matched sub-figures, as expected from the lack of uniqueness of the template. In Table \ref{qatm_table} we present first raw results of said tests.

\begin{table}[H]
    \begin{center}
     \begin{tabular} { c | c | c |  c | c |c |c|c |c|c |c |c  |c|}
     
      {$\textbf{Index}$} & 1 & 2 & 3 & 4 & 5 & 6 & 7 & 8 & 9 & 10 & 11 & 12\\  
      \hline
          $\underset{\textbf{Unique}}{\includegraphics[height=1cm,width=0.75cm]{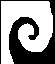}}$  & 
         \includegraphics[height=1.5cm,width=0.75cm]{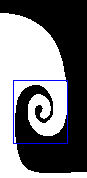} &
          \includegraphics[height=1.5cm,width=0.75cm]{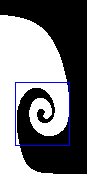} &
          \includegraphics[height=1.5cm,width=0.75cm]{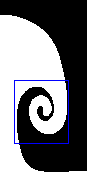} &
          \includegraphics[height=1.5cm,width=0.75cm]{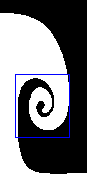} &
          \includegraphics[height=1.5cm,width=0.75cm]{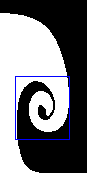} &
          \includegraphics[height=1.5cm,width=0.75cm]{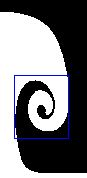} &
          \includegraphics[height=1.5cm,width=0.75cm]{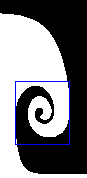} &
          \includegraphics[height=1.5cm,width=0.75cm]{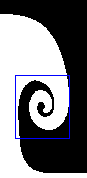} &
          \includegraphics[height=1.5cm,width=0.75cm]{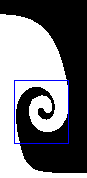} &
          \includegraphics[height=1.5cm,width=0.75cm]{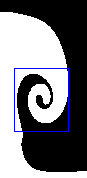} &
          \includegraphics[height=1.5cm,width=0.75cm]{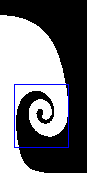} &
          \includegraphics[height=1.5cm,width=0.75cm]{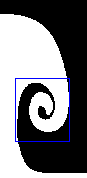}
        \\ \hline
        $\underset{\textbf{Semi-Unique}}{\includegraphics[height=1cm,width=0.75cm]{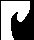}}$  & 
         \includegraphics[height=1.5cm,width=0.75cm]{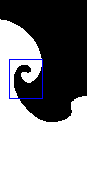} &
          \includegraphics[height=1.5cm,width=0.75cm]{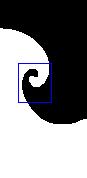} &
          \includegraphics[height=1.5cm,width=0.75cm]{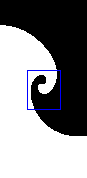} &
          \includegraphics[height=1.5cm,width=0.75cm]{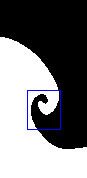} &
          \includegraphics[height=1.5cm,width=0.75cm]{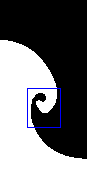} &
          \includegraphics[height=1.5cm,width=0.75cm]{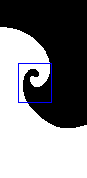} &
          \includegraphics[height=1.5cm,width=0.75cm]{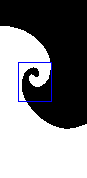} &
          \includegraphics[height=1.5cm,width=0.75cm]{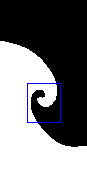} &
          \includegraphics[height=1.5cm,width=0.75cm]{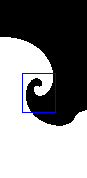} &
          \includegraphics[height=1.5cm,width=0.75cm]{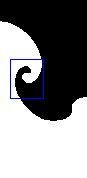} &
          \includegraphics[height=1.5cm,width=0.75cm]{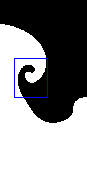} &
          \includegraphics[height=1.5cm,width=0.75cm]{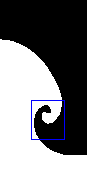}
        \\ \hline
        $\underset{\textbf{Non-Unique}}{\includegraphics[height=1.2cm,width=0.3cm]{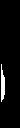}}$ & 
         \includegraphics[height=1.5cm,width=0.75cm]{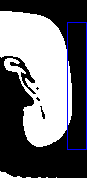} &
         \includegraphics[height=1.5cm,width=0.75cm]{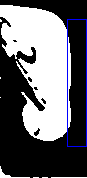} &
         \includegraphics[height=1.5cm,width=0.75cm]{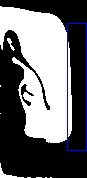} &
         \includegraphics[height=1.5cm,width=0.75cm]{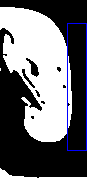} &
         \includegraphics[height=1.5cm,width=0.75cm]{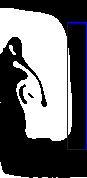} &
         \includegraphics[height=1.5cm,width=0.75cm]{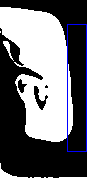} &
         \includegraphics[height=1.5cm,width=0.75cm]{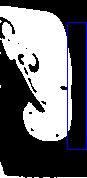} &
         \includegraphics[height=1.5cm,width=0.75cm]{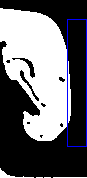} &
          \includegraphics[height=1.5cm,width=0.75cm]{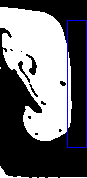} &
          \includegraphics[height=1.5cm,width=0.75cm]{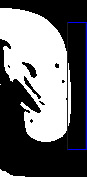} &
          \includegraphics[height=1.5cm,width=0.75cm]{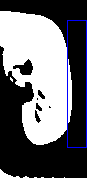}&
          \includegraphics[height=1.5cm,width=0.75cm]{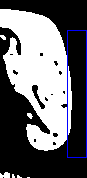}
          \\ \hline 

      \end{tabular}
      \end{center}
      \vspace{-0.5cm}
      \caption{QATM best matchings for the templates discussed  in Fig. \ref{QATMt:all_plots}. }
      \label{qatm_table}
      \end{table}

\subsection{Task \Romannum{4}: Spatiotemporal Prediction}
The PredRNN model was trained on \textit{RayleAI} sequences of 0.01s time steps. As mentioned, the RTI experiment with the parameters $g=740$, estimated amplitude, and $\A=0.155$ contains 12 black and white images with an interval of $\sim$0.033s between each couple of consecutive frames, while the time steps of our simulation are of 0.01s. In order to fill the missing time steps, we used PredRNN to predict the missing time intervals of the experiment. We filled the missing time steps in an iterative manner, by predicting a single future time step at a time. Furthermore, we tested the quality of the prediction of a simulation with the following parameters: $g=725$, $\A=0.14$ and $h=0.3$. As an input, the first 10 images of the simulation were given, while predicting a total of 49 time steps. 

\begin{table}[H]
    \begin{center}
     \begin{tabular} { c | c | c |  c | c |c |c|c |c|c |c |c  |c|}
     
      {$\textbf{Time}$} & 0.03 & 0.06 & 0.1 & 0.13 & 0.16 & 0.2 & 0.23 & 0.26 & 0.3 & 0.33 & 0.36 & 0.4\\  
      \hline
        \textbf{Exp'}  & 
         \includegraphics[height=1.5cm,width=0.75cm]{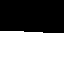} &
          \includegraphics[height=1.5cm,width=0.75cm]{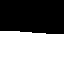} &
          \includegraphics[height=1.5cm,width=0.75cm]{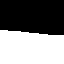} &
          \includegraphics[height=1.5cm,width=0.75cm]{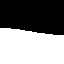} &
          \includegraphics[height=1.5cm,width=0.75cm]{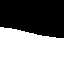} &
          \includegraphics[height=1.5cm,width=0.75cm]{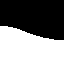} &
          \includegraphics[height=1.5cm,width=0.75cm]{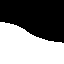} &
          \includegraphics[height=1.5cm,width=0.75cm]{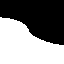} &
          \includegraphics[height=1.5cm,width=0.75cm]{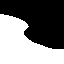} &
          \includegraphics[height=1.5cm,width=0.75cm]{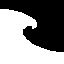} &
          \includegraphics[height=1.5cm,width=0.75cm]{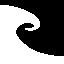} &
          \includegraphics[height=1.5cm,width=0.75cm]{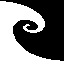}
        \\ \hline

        \textbf{Pred'} & 
         \includegraphics[height=1.5cm,width=0.75cm]{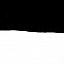} &
         \includegraphics[height=1.5cm,width=0.75cm]{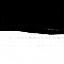} &
         \includegraphics[height=1.5cm,width=0.75cm]{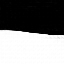} &
         \includegraphics[height=1.5cm,width=0.75cm]{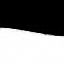} &
         \includegraphics[height=1.5cm,width=0.75cm]{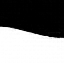} &
         \includegraphics[height=1.5cm,width=0.75cm]{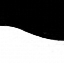} &
         \includegraphics[height=1.5cm,width=0.75cm]{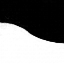} &
         \includegraphics[height=1.5cm,width=0.75cm]{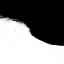} &
          \includegraphics[height=1.5cm,width=0.75cm]{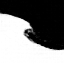} &
          \includegraphics[height=1.5cm,width=0.75cm]{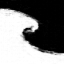} &
          \includegraphics[height=1.5cm,width=0.75cm]{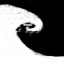}&
          \includegraphics[height=1.5cm,width=0.75cm]{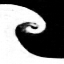}
          \\ \hline 

        \textbf{Sim'} &
          \includegraphics[height=1.5cm,width=0.75cm]{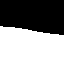} &
          \includegraphics[height=1.5cm,width=0.75cm]{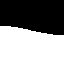} &
          \includegraphics[height=1.5cm,width=0.75cm]{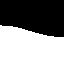} &
          \includegraphics[height=1.5cm,width=0.75cm]{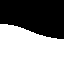} &
          \includegraphics[height=1.5cm,width=0.75cm]{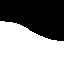} &
          \includegraphics[height=1.5cm,width=0.75cm]{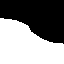}&
          \includegraphics[height=1.5cm,width=0.75cm]{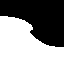} &
          \includegraphics[height=1.5cm,width=0.75cm]{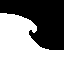} &
          \includegraphics[height=1.5cm,width=0.75cm]{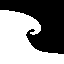} &
          \includegraphics[height=1.5cm,width=0.75cm]{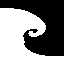} &
          \includegraphics[height=1.5cm,width=0.75cm]{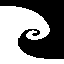} &
          \includegraphics[height=1.5cm,width=0.75cm]{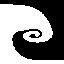}

        \\ \hline
        \textbf{Pred'} &
          \includegraphics[height=1.5cm,width=0.75cm]{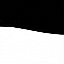} &
          \includegraphics[height=1.5cm,width=0.75cm]{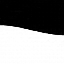} &
          \includegraphics[height=1.5cm,width=0.75cm]{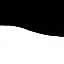} &
          \includegraphics[height=1.5cm,width=0.75cm]{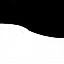} &
          \includegraphics[height=1.5cm,width=0.75cm]{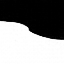} &
          \includegraphics[height=1.5cm,width=0.75cm]{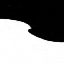} &
          \includegraphics[height=1.5cm,width=0.75cm]{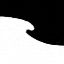} &
          \includegraphics[height=1.5cm,width=0.75cm]{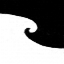} &
          \includegraphics[height=1.5cm,width=0.75cm]{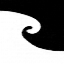} &
          \includegraphics[height=1.5cm,width=0.75cm]{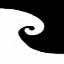} &
          \includegraphics[height=1.5cm,width=0.75cm]{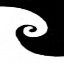} &
          \includegraphics[height=1.5cm,width=0.75cm]{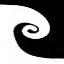}
        \\ \hline
      \end{tabular}
      \end{center}
      \vspace{-0.5cm}
      \caption{PredRNN prediction of the experiment and simulation.}
      \label{predrnn_table}
      \end{table}

\begin{figure}[H]
\centering
\begin{subfigure}[b]{.3\linewidth}
\centering
\includegraphics[width=1\linewidth]{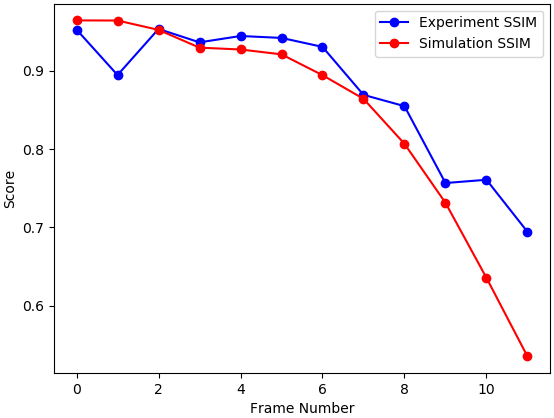}
\caption{\scalebox{0.8}{PredRNN SSIM}} \label{ssim}
\end{subfigure}
\hspace{0.7cm}
\begin{subfigure}[b]{.3\linewidth}
\centering
\hspace{1cm}
\includegraphics[width=1\linewidth]{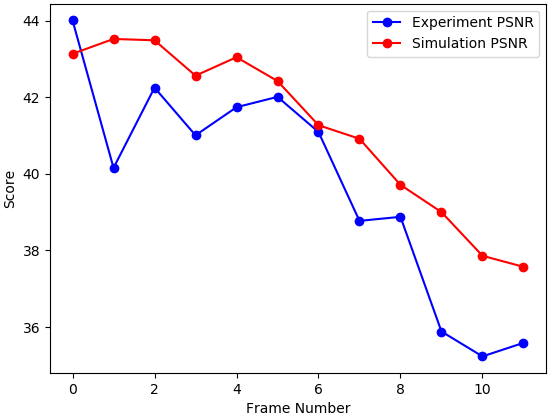}
\caption{\scalebox{0.8}{PredRNN PSNR}} \label{psnr}
\end{subfigure}
\caption{SSIM and PSNR scores of the predicted experimental and simulated results.} \label{psnr_ssim}
\end{figure}

The results of PredRNN prediction are shown in Table~\ref{predrnn_table}. The columns represent time steps ranging from $t_{init}=0.03$ to $t_{fin}=0.4$. The first and third rows of images represent the images of the corresponding time steps of the experiment and the simulation respectively, and as such consider to be GTs. The second and the fourth rows of images represent the prediction of PredRNN on the corresponding time steps of the \textit{filled} experiment and the simulation respectively. As one can see, PredRNN produces very similar predictions to the GT images. 
In Fig.~\ref{psnr_ssim} we quantify the quality of the predicted images using PSNR and SSIM evaluation tests between the produced PredRNN images to their corresponding GT images, similarly to \cite{wang2017predrnn}. Both scores measure the similarity between the predicted frame and its corresponding GT frame. These scores decrease as time progresses, due to the expected difficulty of the model to predict the distant future. However, its worth noting that a simple image sharpening on the predicted results can dramatically increase both SSIM and PSNR scores.

\section{Conclusions and Future Work}
In this work we presented our state-of-the-art complete CVDL methodology for investigating hydrodynamic instabilities. First, we defined the problems and emphasised their significance. Second, we created a new comprehensive tagged database for the needed learning process, which contains simulated diagnostics for training, and experimental ones for testing. Third, we showed how our methodology targets the main acute problems in which CVDL can aid in the current analysis process, namely using deep image retrieval; regressive deep convolutional neural networks; quality aware deep template matching; and deep spatiotemporal prediction. Fourth, we formed a new physical loss and evaluation methodology, which enables to compare the performances of the model against the physical reality, and by such to validate its predictions. At last, we exemplified the usage of the methods on the trained models and assured their performances using our physical loss methodology. In all of the four tasks, we managed to achieve excellent results, which prove the methodology suitability to the problem domain. Thus, we stress that the proposed methodology can and should be an essential part of the hydrodynamic instabilities investigation toolkit, along with analytical models, experiments and simulations. 

In regard to future work, an extension of the methodology might be useful for solving the discrepancy between simulations and experiments when it is clear that the initial parameters of the simulation is not covering the \textit{entire} physical scope. Since in many cases a classical parameter sweep does not yield the desired results, an extension of the model -- in the form of unmodelled parameters regression -- should be used. For example, it might be useful for the physical problem presented in~\cite{kuranz2018high,huntington2018ablative}, in which the simulation results do not cope with the experimental ones. Thus, an extended deep regressive parameter extraction model should be applied in a new form such that unknown parameters -- i.e. parameters which are not part of the simulation initiation -- could be discovered and formulated. This is crucial, as numerous current efforts suggest that often there is a missing part in the understanding of the simulated results. Thus, preventing any traditional method to match the simulated results to the experimental ones. Once discovered, in order to understand and formulate said unknown parameters, an extensive Explainable AI (XAI) methodology \cite{ribeiro2016should} should be performed.

Another strength of the presented methodology is that it can be applied on an already existing data in case that parameter sweep was previously performed on other physical data. Therefore, it might yield physical insights without running any additional simulation. In addition, the toolkit can be easily suited to the physical problem. For example, if the width of vortices was investigated in a previous research \cite{wan2015observation,wan2017observation,shimony2018construction}, template matching would be useful for investigating the inner structure of the vortices; With series of experimental images from different times at hand, the spatiotemporal prediction can be used for prediction of results at unmeasured times. Finally, the models presented in this work might be invaluable for learning physical problems with less training data and a more complex form, using Transfer Learning \cite{tan2018survey}.

\section*{Acknowledgments}
This work was supported by the Lynn and William Frankel Center for Computer Science. Computational support was provided by the NegevHPC project \cite{1negevhpc}.



\bibliographystyle{unsrt} 
\bibliography{bibliography} 

\end{document}